\documentclass{pasa}%

\title[]{Correlated oscillations due to similar multi-path effects seen in two widely separated radio telescopes}
\author[P.N. Diep et al.]{P.N. Diep, N.T. Phuong, P. Darriulat, P.T. Nhung,  P.T. Anh, P.N. Dong, 
D.T. Hoai \and N.T. Thao\\
\affil{VATLY, INST, 179, Hoang Quoc Viet, Cau Giay, Ha Noi, Vietnam}}%
\jid{PASA}
\doi{10.1017/pas.\the\year.xxx}
\jyear{\the\year}

\usepackage{wasysym}
\usepackage{multicol}
\usepackage{float}
\usepackage{graphicx}
\begin{document}%
\begin{abstract}
A multipath mechanism similar to that used in Australia sixty years ago by the Sea-cliff Interferometer is shown to generate correlations between the periods of oscillations observed by two distant radio telescopes pointed to the Sun. The oscillations are the result of interferences between the direct wave detected in the main antenna lobe and its reflection on ground detected in a side lobe. A model is made of such oscillations in the case of two observatories located at equal longitudes and opposite tropical latitudes, respectively in Ha Noi (Viet Nam) and Learmonth (Australia), where similar radio telescopes are operated at 1.4 GHz. Simple specular reflection from ground is found to give a good description of the observed oscillations and to explain correlations that had been previously observed and for which no satisfactory interpretation, instrumental or other, had been found.
\end{abstract}
\maketitle%
\section{INTRODUCTION }
\label{sec:intro}
In the middle of the last century, in the wake of observations made by Navy officers during World War II, according to which the radar echo of a plane flying near horizon above the ocean is modulated by interference fringes, Australia was the home of the founding fathers of radio interferometry and the site of pioneering observations using the so-called ``sea-cliff interferometer'' \cite{1}. The principle of the method was to observe a radio source as it rises above the horizon with a single antenna located on top of a cliff above the ocean; the direct wave and its reflection on the water surface interfere and produce interference fringes that allow for considerably improved angular resolution with respect to what was possible at that time. Observations of solar spots \cite{2}, soon followed by observations of various radio sources \cite{3}, have then been reported.

The present work is an illustration of the same mechanism causing correlations between apparent solar oscillations simultaneously detected by two distant observatories respectively located in Ha Noi (Viet Nam) and Learmonth (Australia) using radio telescopes operated at 1.415 GHz. In this case, oscillations are not observed on the rising Sun but at large elevations: the reflected wave reaches the antenna in one of its side lobes, at large angle with respect to the beam. As a result, the oscillations have amplitudes of a few per mil, rarely exceeding 1\%. They occur on the ground surrounding the antenna and their periods are in the range of a few minutes.

The intriguing existence of correlations between the Ha Noi and Learmonth observations had first been considered as an argument against a possible instrumental effect \cite{4}. It is now clearly established that the cause of the correlation is purely instrumental. The following sections develop this argument, a brief preliminary account of which has been presented elsewhere \cite{5}.

\section{MULTIPATH FROM SPECULAR REFLECTION ON GROUND}
\label{sec:multipath}
Interferences between the direct plane wave emitted by a radio source and detected in an antenna, and its specular reflection on a horizontal surface -- flat ground or ocean -- have been known to produce oscillations since the first days of radio interferometry \cite{1}. Writing $\xi$ the difference in path length between the interfering waves, $\nu$ the frequency and $\lambda$ the wavelength, here $\sim$21 cm, the time difference between the two interfering waves is $\Delta t=\xi/\lambda\nu$. Introducing a parameter $\epsilon$ to account for the attenuation resulting from the reflection on ground and from the lesser gain of the side lobe in which the reflected wave is detected, the dependence on time $t$ of the detected signal reads 
\begin{eqnarray*}
S&&=|e^{i2\pi\nu(t-\frac{1}{2}\Delta t)}+\epsilon e^{i2\pi\nu (t+\frac{1}{2}\Delta t)}|^2\\
&&=|e^{i2\pi\nu t}|^2|e^{–i\pi\xi/\lambda}+\epsilon e^{i\pi\xiξ/\lambda}|^2\\
&&=1+\epsilon^2+2\epsilon cos(2\pi\xi/\lambda).
\end{eqnarray*}

In a time interval centred on $t_0$ (where $\xi=\xi_0$) the time dependence of the oscillation reads therefore
\begin{equation}{\label{eq1}} 
S(t)=1+\epsilon^2+2\epsilon cos\{2\pi\lambda^{-1}[\xi_0+(t-t_0)d\xi/dt]\}
\end{equation}

Describing the oscillation as a sine wave, $sin[2\pi(t-t_0)/T+\varphi]$, implies a period $T=\lambda/(d\xi/dt)$ and a phase $\varphi=\pi/2+2\pi\xi_0/\lambda$. In order to have $T$ positive, one prefers to write
\begin{equation}{\label{eq2}} 
T=\lambda/|d\xi/dt| \text{ and } \varphi=\pi/2\pm2\pi\xi_0/\lambda 			    			      
\end{equation}

where the $+$ sign is for $\xi$ increasing with $t$ and the $-$ sign for $\xi$ decreasing with $t$. Relation 2 implies a fundamental relation between $T$  and $d\varphi/dt$
\begin{equation}{\label{eq3}}
T|d\varphi/dt|=2\pi
\end{equation}	
which is characteristic of the multipathing effect.
 
Note that Relation 2 allows for an evaluation of $d\xi/dt$ from measurements of $T$ and/or $d\varphi/dt$ but not for a measurement of $\xi$ itself: the difference in path length can only be obtained up to an unknown integer number of wavelengths. Equivalently, $\varphi$ is only defined up to an unknown integer factor of $2\pi$.
 
From Figure 1, $\xi=2Dsin(h)$ where $D$ is the distance of the antenna to ground and $h$ the elevation of the source.

As $h$ increases with $t$ in the morning and decreases with $t$ in the afternoon, so does $\xi$. At a same time $t$, from one day to the next, one expects therefore $\varphi$ to increase in the morning in the Northern hemisphere between December 22nd and June 21st. Changing from morning to afternoon, or from Northern to Southern hemisphere or from December-June to June-December changes the sign of the daily variation of $\varphi$.

If the hypothesis of specular reflection is slightly relaxed, reflections with an impact point farther away from the antenna than the specular impact (i.e. reaching the dish at a smaller angle with respect to the main lobe) are favoured with respect to reflections with an impact point closer than the specular impact. The reason is not only from pure geometry arguments but also because the reflected wave reaches the dish at large angle with respect to the beam and the antenna gain, on average, decreases with this angle. This causes the mean impact to be behind the specular impact and the path difference $\xi$ to take larger values than in the specular case; as $D$ is proportional to $\xi$, it is overestimated accordingly.
 
The effect is easily simulated by describing the deviation from specular reflection by a Gaussian dependence of the reflection probability on the angle $r$ of the reflected wave with respect to the specularly reflected wave, $exp(-\frac{1}{2}r^2/\sigma_r^2)$ and the decrease of the antenna gain by another Gaussian dependence on the angle $\theta$ between the reflected wave and the antenna axis, $exp(-\frac{1}{2}\theta^2/\sigma_\theta^2)$ (Figure 1). Moreover, accounting for the irregularities of the reflecting surface can be approximated by a smearing of $\xi$ using a third Gaussian,  $exp[-\frac{1}{2}\Delta\xi^2/(\xi\sigma_{sm})^2]$. 

\begin{figure}[ht]
\begin{center}
\includegraphics[width=0.38\textwidth]{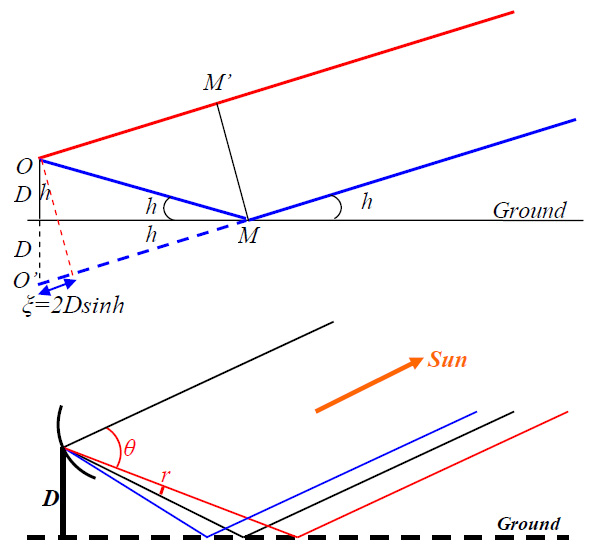}
\caption{Upper panel: geometry of specular reflection on ground into a dish centred in $O$ and having image $O'$ in the ground mirror. Lower panel: departure from exact specular reflection (mean ray), definition of the angles $r$ and $\theta$.}\label{Fig1}
\end{center}
\end{figure}

An important consequence of the above mechanism is the existence of a correlation between the periods of simultaneous oscillations independently detected by two distant observatories such as Ha Noi and Learmonth.
 
Consider two radio telescopes operated at a same frequency at heights $D_1$ and $D_2$ above a flat ground in observatories located at respective longitudes and latitudes $(\psi_1,\zeta_1)$ and $(\psi_2,\zeta_2)$.
 
Writing $h'=|dh/dt|$, the periods read $T_{1,2}=\frac{1}{2}\lambda/[D_{1,2}h’_{1,2}cos(h_{1,2})]$. In the approximation of a circular Earth orbit, calling $\delta$ the Sun declination and $H=t+\psi$ the hour angle $sin(h)=sin\delta sin\zeta+cos\delta cos\zeta cosH$ ($H=0$ when the Sun crosses the meridian plane). Hence,
\begin{equation} 
T=T^*/[cos\delta|sin(t+\psi)|]
\end{equation}

with $T^*=\frac{1}{2}\lambda/(Dcos\zeta)$ being a constant for each observatory. Writing $\rho=T/(T^*cos\delta)$, $\rho_1=1/|sin(t+\psi_1)|$ and $\rho_2=1/|sin(t+\psi_2)|$ are trivially correlated, implying a correlation between the periods measured simultaneously by both observatories. If the two observatories are exactly at the same longitude, $\rho_1=\rho_2$ at any time. In practice, the longitudes of the two observatories should not be too different for them to have a chance to observe the Sun simultaneously for long periods of time. In the Ha Noi/Learmonth case, the two latitudes are nearly opposite ($\pm 21.6^\circ\pm 0.6^\circ$) and the two longitudes are nearly equal ($110^\circ {\text{E}}\pm 4^\circ$). In the approximation where $\psi_1=\psi_0-\Delta$, $\psi_2=\psi_0+\Delta$, $0<\Delta<<1$ rad (observatory 1 being east of observatory 2) and writing $H_0=\psi_0+t$ the mean hour angle, 
\begin{equation}
\rho_1=1/|sin(H_0+\Delta)| \text{ and } \rho_2=1/|sin(H_0-\Delta)|
\end{equation}

\begin{figure}[ht]
\begin{center}
\hspace{0.55cm}\includegraphics[width=0.275\textwidth]{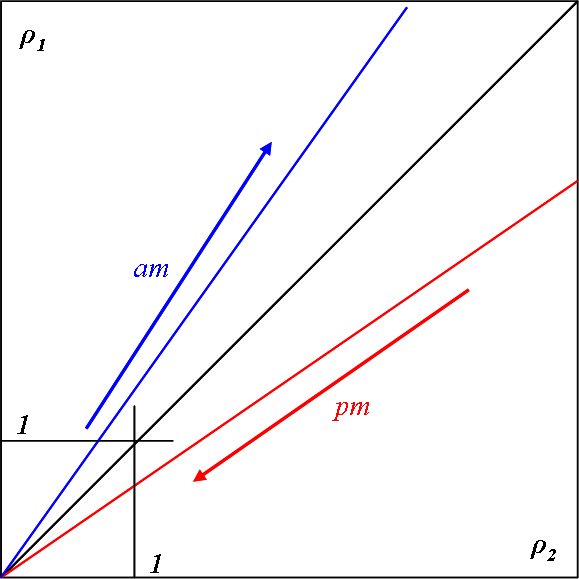}
\includegraphics[width=0.33\textwidth]{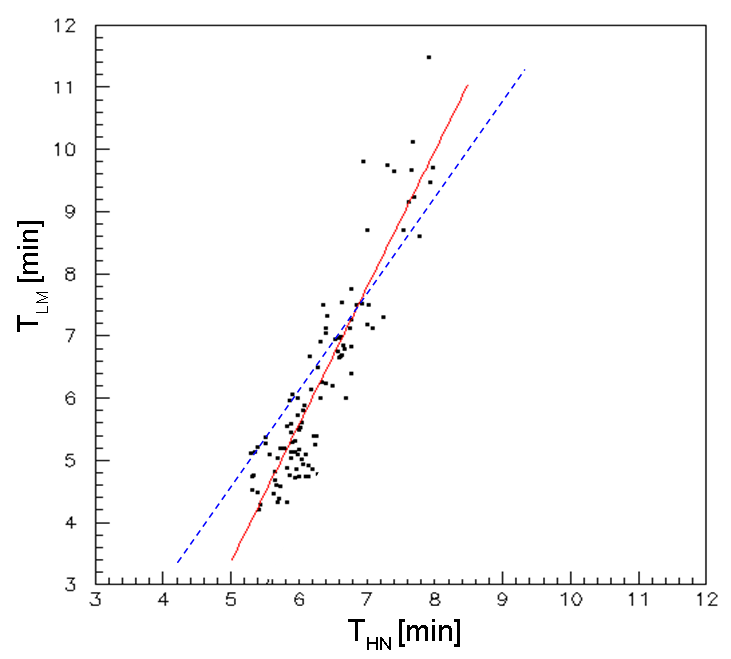}
\caption{Correlations observed between the periods of oscillations measured in two observatories at nearby longitudes. Upper panel: schematic illustration of the main features; the angle between the morning and afternoon lines is a measure of the difference of longitude between the two observatories. Lower panel: correlation observed in Reference 1 between Learmonth (ordinate) and Ha Noi (abscissa); the dotted line displays the model prediction for morning oscillations using respective $D$ values of 7 m and 6 m for Learmonth and Ha Noi respectively.}\label{Fig2}
\end{center}
\end{figure}

Around noon, the elevation is stationary and the period of the oscillations takes very large values outside the range where they can be detected. In the morning, $H_0<-\Delta$, the elevation, and therefore the period of the oscillations increases with time. In the afternoon, $H_0>\Delta$, they decrease. Changing $H_0$ into $-H_0$ changes $\rho_1$ into $\rho_2$. In the morning, both periods increase, that of observatory 1 faster than that of observatory 2. In the afternoon, both decrease, that of observatory 2 faster than that of observatory 1. This is schematically illustrated in the upper panel of Figure 2. 	
The predicted correlation is essentially independent of the season during which observations are made: the only seasonal contribution is from the $cos\delta$ term which remains confined between 0.92 and 1 over the whole year and has therefore very little effect.

\section{OBSERVED OSCILLATIONS IN LEARMONTH AND HANOI}
\begin{figure}[ht]
\begin{center}
\includegraphics[width=0.45\textwidth]{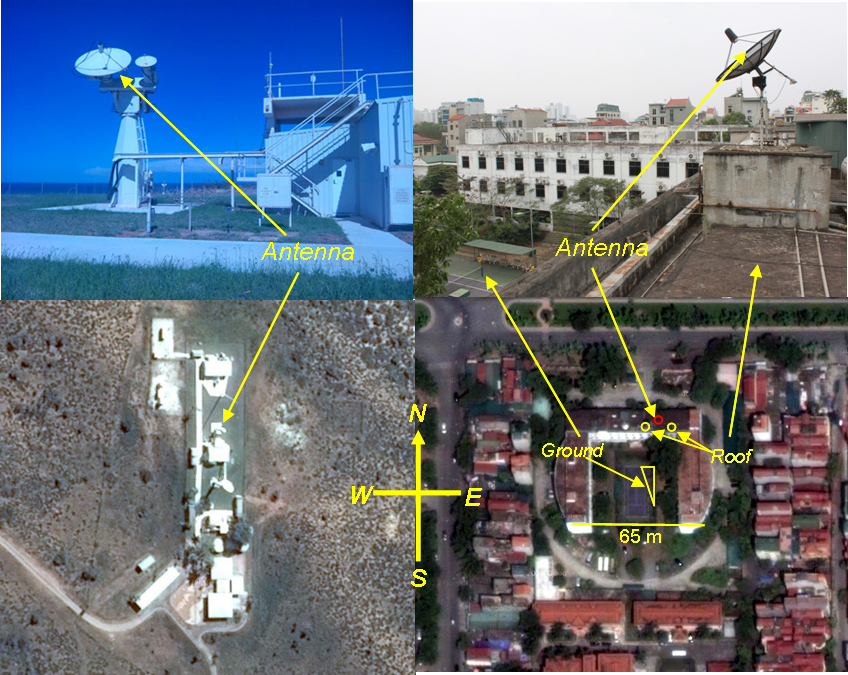}
\caption{Sites of the observatories in Learmonth (left, courtesy of Dr Owen Giersch) and Ha Noi (right). The lower panels show satellite maps of the two sites (source: Google map).}\label{Fig3}
\end{center}
\end{figure}

When applied to a real observatory, the parameters that characterize the oscillations are strongly dependent on the environment of the antenna. In the cases studied here, Ha Noi and Learmonth, the distance of the antenna to the reflecting surface is between 5 and 20 m, the latitude is $\sim\pm22^\circ$ and the wavelength is 21 cm. Hence $T^*$ is between 1 and 5 min. The method used to detect oscillations requires their period to be of the order of a few minutes, say 2 to 10 minutes typically, excluding an interval of 1 to 2 hours around local noon, during which the period of the oscillations is too large to be observed. However, the rest of the time, one deals with a broad range of elevations, allowing the impact point on ground to reach close to the antenna.

The Learmonth antenna, 2.4 m (8 ft) in diameter, (Figure 3 left) is located in an airport on a pole erected some 7.5 m above a flat ground. The only nearby building is located south of the antenna, outside the range of azimuths toward which the antenna is pointing. On the contrary, the Ha Noi antenna, 2.6 m in diameter, (Figure 3 right) is on top of a small building, some 20 m above ground and some 5 m above the flat roof, in a urban environment. One would therefore expect the multipath pattern observed at Learmonth to be considerably simpler and easier to describe than that observed in Ha Noi. 
 
The data used in the present work cover the period October 24th 2013 to January 31st 2014 for both Ha Noi and Learmonth. In addition Learmonth data collected during the 10 central days of each month of the year have been included in the analysis. For both Ha Noi and Learmonth, the signal to noise ratio is of the order of 500. Details of the procedure of data reduction are available in References 5 and 6. 
 
After subtraction of solar flares and occasional interferences of human origin, the time dependence of the antenna temperature, averaged over the bandwidth, is searched for oscillations having periods in the 2 to 10 minutes interval. The method consists in evaluating, for each independent measurement, the amount of oscillation that the data can accommodate. Each oscillation is evaluated over a 20 min interval centred on the particular measurement $i$ and is characterized by an amplitude $A_i$, a phase $\varphi_i$, a period $T_i$ and a chi square describing the quality of the fit, $\chi^2_i$.  The significance of such oscillations can then be assessed objectively from the values of $A_i$ and $\chi^2_i$ without requiring human intervention. For each period $T$ and each measurement $i$ one calculates the mean value $M_i$ and the rms deviation with respect to the mean $R_i$ in a time interval centred on measurement $i$ and having a width $T$. In a second step, a fit of the antenna temperature is performed over a 20 min time interval centred on measurement $i$ to a form $M_i+\eta R_isin[2\pi(t-t_i)/T+\varphi]$. Here $t$ spans the 20 min interval while $t_i$ is the time of measurement $i$. Parameters $\eta$ and $\varphi$ measure the relative amplitude and phase of the oscillation respectively. The problem being linear in $\eta cos\varphi$ and $\eta sin\varphi$ allows for an easy explicit calculation of the parameters. Figure 4 illustrates the procedure.

Significant oscillations are defined as having a small $\chi^2_i$ value, a large $\eta$ value (meaning that the oscillation accounts for a major fraction of the signal fluctuation in the 20 min interval) and a large $\eta R_i/M_i$ value (meaning that the amplitude of the oscillation exceeds noise level). Different selection criteria have been tried and the robustness of the corresponding conclusions has been ascertained.

Distributions of time versus period using sensible selection criteria display very clear patterns that are illustrated in Figure 5 and follow the same trend as expected from specular reflection on ground. Examples of the evolution of the phase of the oscillations from one day to the next are displayed in Figure 6. Here again, the dependence on season, hemisphere and time in the day is as expected from specular reflection on ground.

\begin{figure}[ht]
\begin{center}
\includegraphics[width=0.35\textwidth]{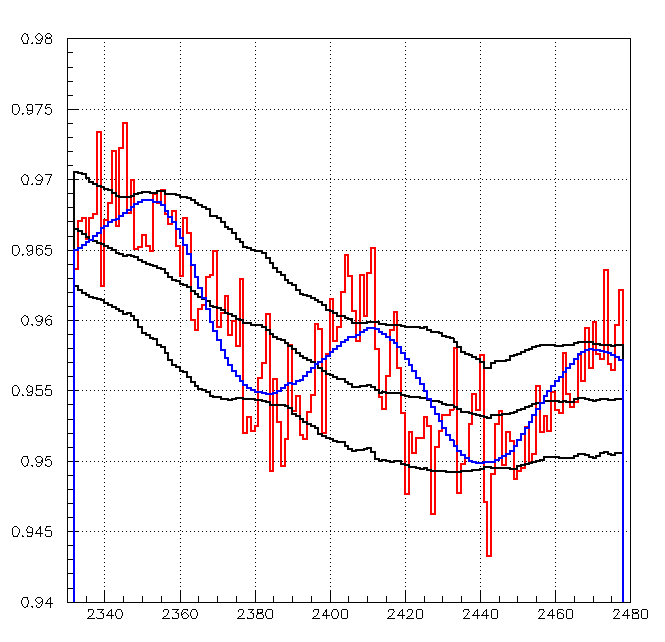}
\hspace{0.5cm}\includegraphics[width=0.325\textwidth]{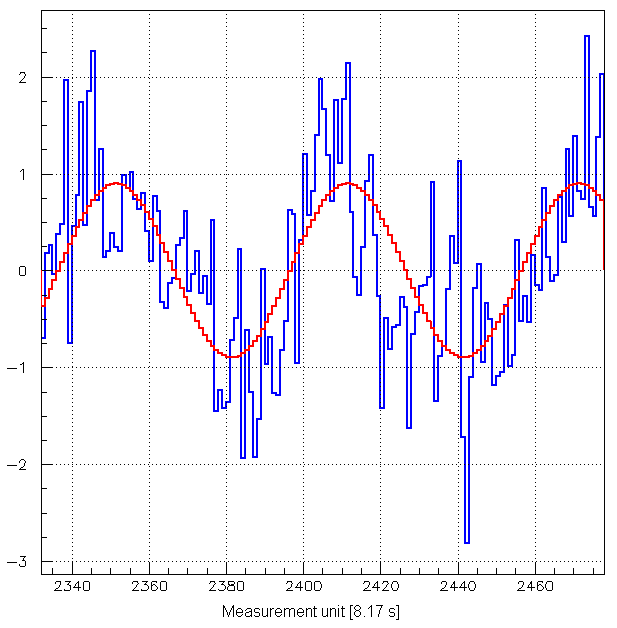}
\caption{A typical oscillation. The upper panel shows the data (red) together with the fit (blue) and $M$ and $M\pm R$ (black). The lower panel compares data (blue) and fit (red) after subtraction of $M$ and division by $R$.}\label{Fig4}
\end{center}
\end{figure}

\begin{figure}[ht]
\begin{center}
\includegraphics[width=0.3\textwidth]{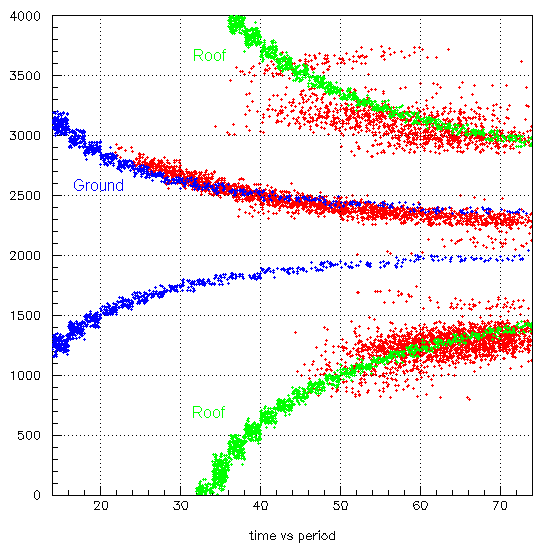}
\includegraphics[width=0.3\textwidth]{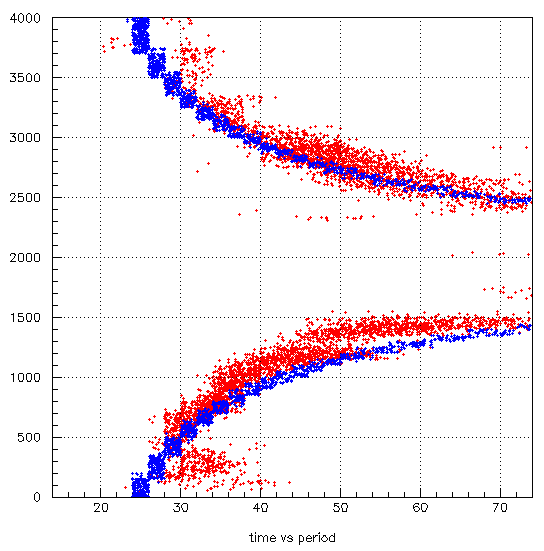}
\caption{Time versus period scatter-plots. Upper panel: Ha Noi data (red) collected between October 25th and December 17th, 2013 . The lines are specular reflection multipath predictions for $D=6$ m (roof) and $D=25$ m (ground). Lower panel: Learmonth data (red) collected in the 10 central days of May 2012. The blue lines are ground specular reflection multipath predictions for $D=8.5$ m.}\label{Fig5}
\end{center}
\end{figure}

\section{COMPARISON BETWEEN OBSERVATIONS AND PREDICTIONS}
Having illustrated by a few examples in Figures 5 and 6 the good qualitative agreement between the observed oscillations and the predictions of a multipath model assuming perfect specular reflection on ground, we now attempt a more quantitative analysis of the effect. The measurements of the period and phase of the oscillations provide independent evaluations of the altitude $D$ of the antenna above ground. From $T=\lambda/|d\delta/dt|=\frac{1}{2}\lambda/(D|dh/dt|cos(h))$ one obtains
\begin{equation}
D=\frac{1}{2}\lambda/(T|dh/dt|cos(h))
\end{equation}

\begin{figure}[ht]
\begin{center}
\includegraphics[width=0.48\textwidth]{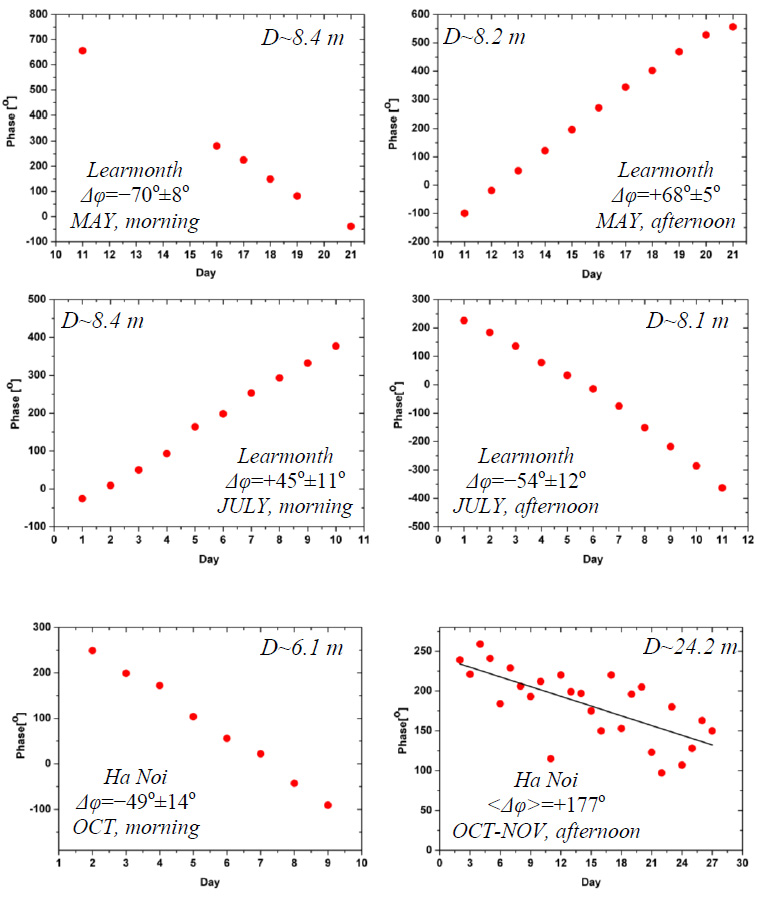}
\caption{Dependence on the date of the phases of oscillations observed under different conditions. The lower right panel displays the daily phase increment rather than the phase itself and is seen to decrease when approaching the Winter solstice as expected (its large value results from the large associated $D$ value).}\label{Fig6}
\end{center}
\end{figure}

Relation 6 relates the height $D$ of the antenna above the reflective surface to the measured value of the period of the oscillations under the hypothesis of specular reflection. Figure 7 displays the distributions of $D$ obtained this way for the oscillations observed in Learmonth and Ha Noi. The former is dominated by ground reflections while the latter displays a more complex structure revealing reflections from the observatory roof in the morning and late afternoon and from ground in the early afternoon. Knowing $D$, it is easy to map the impact coordinates on ground, $x=Dcos(a)/tan(h)$ and $y=Dsin(a)/tan(h)$. They do not display any particular structure in the Learmonth case, which is consistent with a flat ground, while revealing clearly distinct regions in the Ha Noi case, that are unambiguously associated with the topography of the environment as shown in Figure 3.
  
Using reasonable values for the parameters describing a small departure from exact specular reflection on ground, it is easy to obtain a good description of the $D$ distributions as illustrated in Figure 7 where a common value of $45^\circ$ has been used for $\sigma_\theta$ while $\sigma_r$  takes values between $9^\circ$ and $16^\circ$ and $\sigma_{sm}\sim 10\%$. The quality of the data does not allow to measure these parameters precisely and other combinations of their values can be found that give also acceptable results. However, in all cases, the departure from exact specular reflection that can be accommodated remains small, not exceeding $20^\circ$. An effect of the inclusion of such departure is to significantly lower the estimate of $D$ with respect to exact specular reflection: from 8.3 m to 7.7 m in the case of Learmonth and, in the case of Ha Noi, from 6.2 m to 5.7 m for the roof and from 26 m to 21 m for ground, improving significantly the agreement with real dimensions (respectively 7.5 m, 5.6 m and 17.7 m). 

\begin{figure}[ht]
\begin{center}
\includegraphics[width=0.3\textwidth]{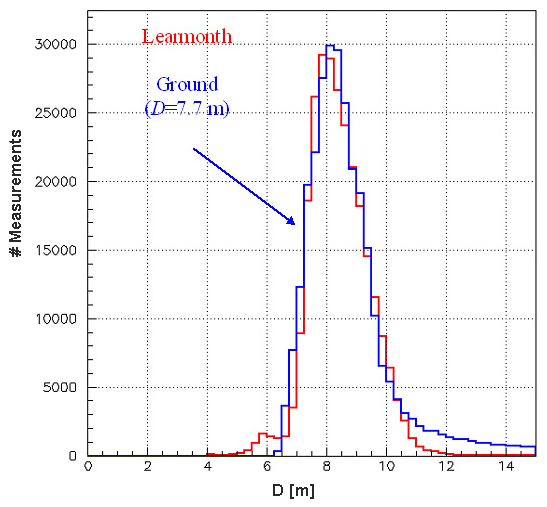}
\includegraphics[width=0.3\textwidth]{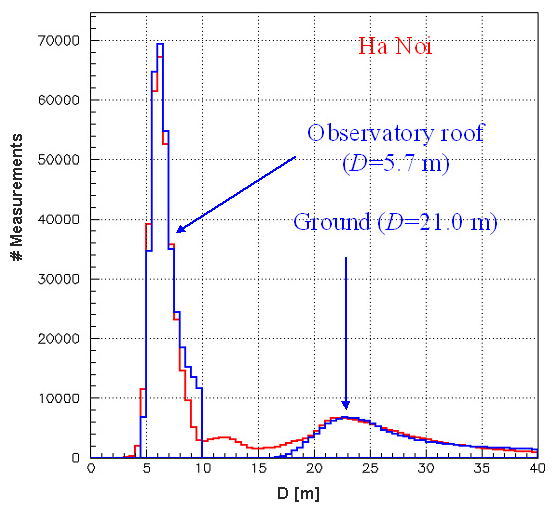}
\caption{$D$ distributions obtained from the Learmonth (upper panel) and Ha Noi (lower panel) data in November-December 2013. The blue lines show model predictions allowing for small departures from exact specular reflections (see text). }\label{Fig7}
\end{center}
\end{figure}

The correlation expected between oscillations observed in the morning at Learmonth from ground and in Ha Noi from reflections on the observatory roof is illustrated in Figure 2 (lower panel). The agreement with observation is remarkable given the crudeness of the model used.

\begin{figure}[ht]
\begin{center}
\includegraphics[width=0.3\textwidth]{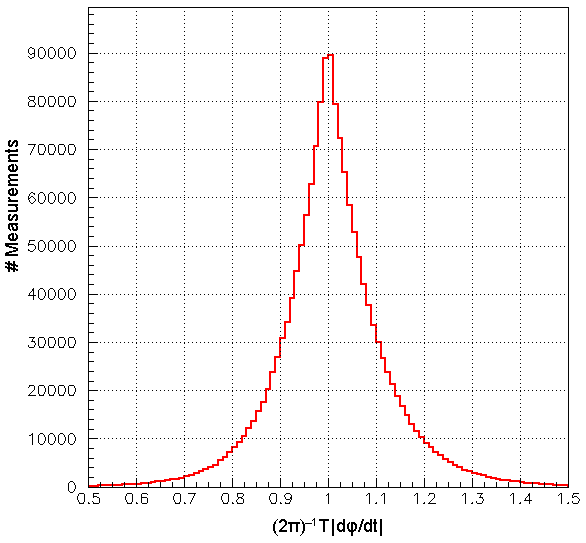}
\includegraphics[width=0.3\textwidth]{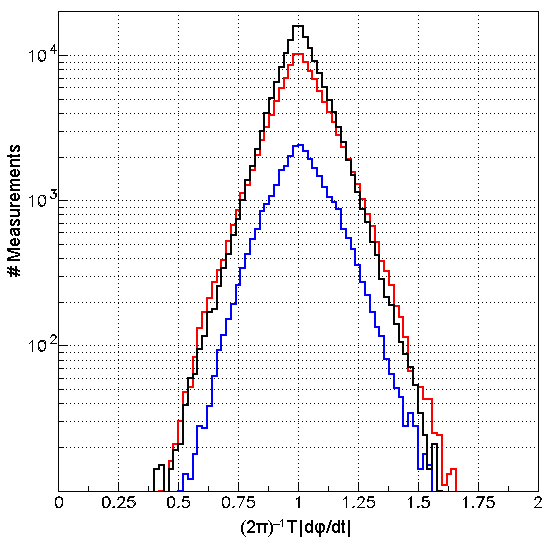}
\caption{Distribution of $(2\pi)^{-1}T|d\varphi/dt|$ for oscillations having amplitudes in excess of $3\permil$ for Learmonth (upper panel) and Ha Noi (lower panel) data. The Ha Noi distributions display separately ground reflections (black) and roof reflections (blue in the morning and red in the afternoon). The scale of ordinate is linear for Learmonth and logarithmic for Ha Noi.}\label{Fig8}
\end{center}
\end{figure}

Relation 3, which relates independent measurements of the period and phase of the observed oscillations, offers a crucial test of their multipath nature. Note that the path difference between the direct and reflected waves is of the order of magnitude of $D$ for the rather large values of the Sun elevation associated with the observed oscillations, meaning several tens of wavelengths. As remarked earlier, the phase of the oscillation is measured up to an integer multiple of $2\pi$ and it is not possible to measure $D$ directly from a single phase value but $d\varphi/dt$ is simply evaluated from the phase difference between two successive measurements. The distribution of $(2\pi)^{-1}T|d\varphi/dt|$ obtained this way is displayed in Figure 8 for Learmonth and Ha Noi oscillations having amplitudes in excess of $3\permil$, giving in both cases very strong evidence for the multipath nature of the effect. Mean(rms) values of 1.01(0.11) and 1.00(0.13) are obtained for Learmonth and Ha Noi respectively, giving strong evidence for the multipath origin of the observed oscillations. Note that Ha Noi data mix reflections from ground and from the observatory roof, spanning a broad range of $D$ values. The Learmonth data cover the whole year and the simple topography implies reflections from a flat ground with a well defined $D$ value, 8.24 m on average. It fluctuates by only $\pm 0.37$ m over the year and the rms value of its distribution is 1.21 m on average. Comparing the summer months (November to February) with the winter months (May to August) for the retained oscillations, the mean elevation of the Sun is found to vary from $49^\circ$ to $35^\circ$ and the mean amplitude of the oscillations from $5.0\permil$ to $6.8\permil$ while the width of the distribution of $(2\pi)^{-1}T|d\varphi/dt|$ remains constant to better than 10\% of its value. The number of retained oscillations is nearly twice as large in winter than in summer.
 
\section{CONCLUSION}
When observing the Sun, multipath effects between the direct wave reaching the antenna in the main lobe and its reflection on ground reaching the antenna in a side lobe have been shown to produce correlations between the periods of oscillations observed independently by two distant radio telescopes. The case of observations made at 1.4 GHz in Ha Noi (Viet Nam) and Learmonth (Australia) has been studied in some detail. Strong evidence for the multipath origin of the observed oscillations has been obtained from the relation between their periods and their phases. Good agreement between observations and model predictions has been obtained and the departure from exact specular reflection that the data can accommodate has been shown to be small. The oscillations have periods and phases that are remarkably simple functions of time and are well described by the model. Their amplitudes, at the level of a few per mil, are consistent with the gain drop expected between the main and side lobes of the antenna pattern. Indeed, for a 65\% aperture efficiency \cite{6} we expect a gain of some 30 dBi for the main lobe compared to some $-$20 dBi for a typical side lobe, namely a voltage ratio of some 3$\permil$. The existence of a correlation between independent observations from two distant observatories, together with the large values of the elevation at which the oscillations were observed, had been used earlier \cite{4} as arguments against an instrumental explanation. It is now clear that the effect is of purely instrumental nature, making a search for genuine solar oscillations in this range of periods and amplitudes unfeasible with such instruments.

\begin{acknowledgements}
We are deeply indebted to the Learmonth Solar Observatory staff, who are making their data available to the public, and particularly to Dr Owen Giersch for having kindly and patiently answered many of our questions related to such data and, in particular, for having first mentioned a possible contribution of multipathing. We are grateful to Dr Alain Maestrini, Dr Pierre Lesaffre and Dr Alan Rogers and the anonymous referee for very useful comments. We acknowledge financial support from the Vietnam National Foundation for Science and Technology Development (NAFOSTED) under grant number 103.08-2012.34, the Institute for Nuclear Science and Technology, the World Laboratory, the Odon Vallet Foundation and the Rencontres du Vietnam.
\end{acknowledgements}


\end{document}